\documentclass[preprint,12pt]{elsarticle}

\usepackage{graphicx}
\usepackage{amssymb}
\usepackage{amsthm}
\usepackage{mathtools}
\usepackage{lineno}
\usepackage{mathrsfs}
\usepackage{booktabs, caption, makecell}
\usepackage{threeparttable}
\usepackage{algorithm}
\usepackage{algcompatible}
\usepackage{lipsum}
\usepackage{appendix}
\usepackage{float}

\usepackage[colorlinks,linkcolor=blue]{hyperref}
\usepackage[nameinlink,capitalise]{cleveref}
\usepackage{calc}

\makeatletter
\newcommand*{\rom}[1]{\expandafter\@slowromancap\romannumeral #1@}
\makeatother

\theoremstyle{remark}
\newtheorem*{remark}{Remark}

\theoremstyle{definition}

\journal{Structural Safety}

\begin{document}

\begin{frontmatter}
\title{Probabilistic Evolution of Stochastic Dynamical Systems: A Meso-scale Perspective}
\author[label1]{Chao Yin}
\author[label1]{Xihaier Luo}
\ead{xluo1@nd.edu}
\author[label1]{Ahsan Kareem \corref{cor1}}
\ead{kareem@nd.edu}
\cortext[cor1]{Corresponding author. 156 Fitzpatrick Hall, Notre Dame, IN 46556, USA.}
\address[label1]{NatHaz Modeling Laboratory, University of Notre Dame,  Notre Dame, IN 46556, United States}

\begin{abstract}
    Stochastic dynamical systems arise naturally across nearly all areas of science and engineering. Typically, a dynamical system model is based on some prior knowledge about the underlying dynamics of interest in which probabilistic features are used to quantify and propagate uncertainties associated with the initial conditions, external excitations, etc. From a probabilistic modeling standing point, two broad classes of methods exist, i.e. macro-scale methods and micro-scale methods. Classically, macro-scale methods such as statistical moments-based strategies are usually too coarse to capture the multi-mode shape or tails of a non-Gaussian distribution. Micro-scale methods such as random samples-based approaches, on the other hand, become computationally very challenging in dealing with high-dimensional stochastic systems. In view of these potential limitations, a meso-scale scheme is proposed here that utilizes a meso-scale statistical structure to describe the dynamical evolution from a probabilistic perspective. The significance of this statistical structure is two-fold. First, it can be tailored to any arbitrary random space. Second, it not only maintains the probability evolution around sample trajectories but also requires fewer meso-scale components than the micro-scale samples. To demonstrate the efficacy of the proposed meso-scale scheme, a set of examples of increasing complexity are provided. Connections to the benchmark stochastic models as conservative and Markov models along with practical implementation guidelines are presented.
\end{abstract}


\end{frontmatter}

\section{Introduction}
\label{sec1}
To accurately capture the dynamical behavior of dynamical systems, it is essential for one to assess the effects of the input uncertainties on model predictions \cite{der2009aleatory, helton2011quantification}. To introduce the methodology, consider a continuous dynamical system evolving on a smooth manifold: 

\begin{equation}
    \label{eq: sec1_1}
    \frac{\mathrm{d} \boldsymbol{x}(t)}{\mathrm{d} t} = f_{\boldsymbol{\theta}}(\boldsymbol{x}, t)
\end{equation}

where the underlying dynamical system $\boldsymbol{x}$ evolves in a complete metric space with a countable dense set $\boldsymbol{\mathcal{X}}$, vector $\boldsymbol{\theta} \in \boldsymbol{\Theta}$ defines the model parameters, $t$ is the temporal index, and $f(\cdot)$ is a Lipschitz vector field \cite{strogatz2001nonlinear}.

From a dynamical evolution perspective, the state $\boldsymbol{x}_{t+1}$ is uniquely determined by the state $\boldsymbol{x}_{t}$ and possibly some noise following the ergodic theory \cite{giannakis2019data}. The discrete-time dynamics can be defined as: 

\begin{equation}
    \label{eq: sec1_2}
    \boldsymbol{x}_{k+1} = \mathbf{F}\left(\boldsymbol{x}_{k}\right) = \boldsymbol{x}_{k}+\int_{k \Delta t}^{(k+1) \Delta t} f(\boldsymbol{x}(\tau)) d \tau
\end{equation}

where $\mathbf{F}(\cdot): \boldsymbol{\mathcal{X}} \rightarrow \boldsymbol{\mathcal{X}}$ is a smooth diffeomorphism \footnote{An invertible function that maps one differentiable manifold to another in such a way that both the function and its inverse are smooth.} that maps the state $\boldsymbol{x}_{k}$ to $\boldsymbol{x}_{k+1}$. In this case, all uncertainty in the system originates from the uncertainty in the initial system state $\boldsymbol{x}(t=0)$ \cite{giannakis2019data, arbabi2017ergodic}. 

This study assumes that \cref{eq: sec1_2} is configured with a random initial state with known joint probability density function $p(\boldsymbol{x}_0)$. The goal here is to carry out uncertainty quantification and propagation by means of probability density function (PDF). Therefore, the solution scheme requires the reformulation of the governing equations from ordinary/partial differential equations (ODEs/PDEs) to its stochastic format \cite{li2009stochastic, ghanem2003stochastic}. Several approaches have been developed in the past few decades, which can be classified into two main categories: macro-scale methods and micro-scale methods.

\subsection{Macro-scale evolution}
\label{sec11}
Macro-scale methods focus on the computation of the distribution parameters, for instance, mean $\mu$ and standard deviation $\sigma$ in the case of normal distribution \cite{melchers2018structural, low2013new}. In practice, an analytical expression in terms of a general probability evolution equation is available only for limited linear systems, where the task of tracking the evolution of the PDF can be accomplished using homogeneity and additivity properties. Specifically, if the input PDF is Gaussian, the evolution of probability can be hereby represented by a sequence of Gaussian density functions with a shifted center and scaled variation. Similar analytical expressions for nonlinear systems may be obtained via linearization, whereas these methods are very effective if only the second-order statistics of the probability evolution is of interest. For a more general and highly nonlinear system, macro-scale methods have difficulty in describing the multi-mode shape or tails of a non-Gaussian distribution \cite{cook2016gaussian, kareem2008numerical}. 

\subsection{Micro-scale trajectory}
\label{sec12}
Micro-scale methods resort to a set of random realizations of the uncertain inputs using Monte Carlo (MC) sampling, or collocation points. In particular, collocation based methods use a grid of points for which PDFs are obtained through finite difference (FD) or finite element methods (FEM) \cite{ghanem2003stochastic, melchers2018structural, xiu2002wiener}. The global PDF then results from a mixture of these PDFs. Interpolation functions are used to describe the PDF of any arbitrary point in space and time, which allows the description of the probability densities at any arbitrary points in addition to the interpolation ones. However, it should be noted that FD and FEM may encounter challenges when applied to high-dimensional random space since their integral domain is not tailored to the random space. By contrast, MC methods use the ensemble of independent sample trajectories to represent the evolution of probability, where the PDF at any time is approximated by a set of discrete probabilities at sample points \cite{robert2013monte}. MC therefore disregards information given by the PDF around particular samples, or, in other words, it assumes the region of a sample to be homogeneous. Therefore, in order to minimize the error in the representation of the real PDF, MC has to employ a large number of samples to fill the random space, or to keep the region of each sample as small as possible. Considering the computational cost of examining a great number of samples, possible remedy strategies include (a) constructing an easy-to-evaluate  surrogate/emulator that is trained using a small number of propagated data \cite{rajashekhar1993new, luo2020bayesian, luo2019deep}; (b) utilizing advanced sampling strategies such as importance sampling, adaptive sampling, etc, to generate high-quality samples \cite{nie2004new, au2001estimation, papaioannou2015mcmc}; and (c) performing the prescreening of the most influential variables by global sensitivity analysis techniques \cite{sudret2008global, wei2015variable}.

\subsection{Meso-scale parcel}
\label{sec13}
In this study, a generalized perspective regarding the investigation of the dynamical evolution of the PDF of interest is introduced following the seminal works presented in \cite{Yin2013theory, Yin2013method}. Viewed from this perspective, the PDF of the input uncertainties are modeled by a mixture of Gaussians \cite{bishop2006pattern, kurtz2013cross}. Characterization of each component of the mixture includes the computation of the mean and covariance, and an active learning method is hereby proposed to accelerate the parameter estimation, where the means of these Gaussian components are directly estimated by a low-discrepancy sequence or a clustering algorithm \cite{robert2013monte, zhang2013structural, kaufman2009finding}, and the optimal covariances are obtained through a complexity-reduced expectation maximization (EM) algorithm \cite{dempster1977maximum, mclachlan2007algorithm}. Next, the evolution of each Gaussian component is expressed in a convolution format that produces an evolutionary kernel \cite{li2004probability, wan2000unscented}. The integral can be efficiently solved by the third-degree spherical-radial cubature rule \cite{arasaratnam2009cubature, jia2013high}. Finally, by assembling the resolved integral results of each component provides the evolutionary PDF of interest. 

\begin{figure}[b!]
    \centering
    \includegraphics[width=1.0\textwidth]{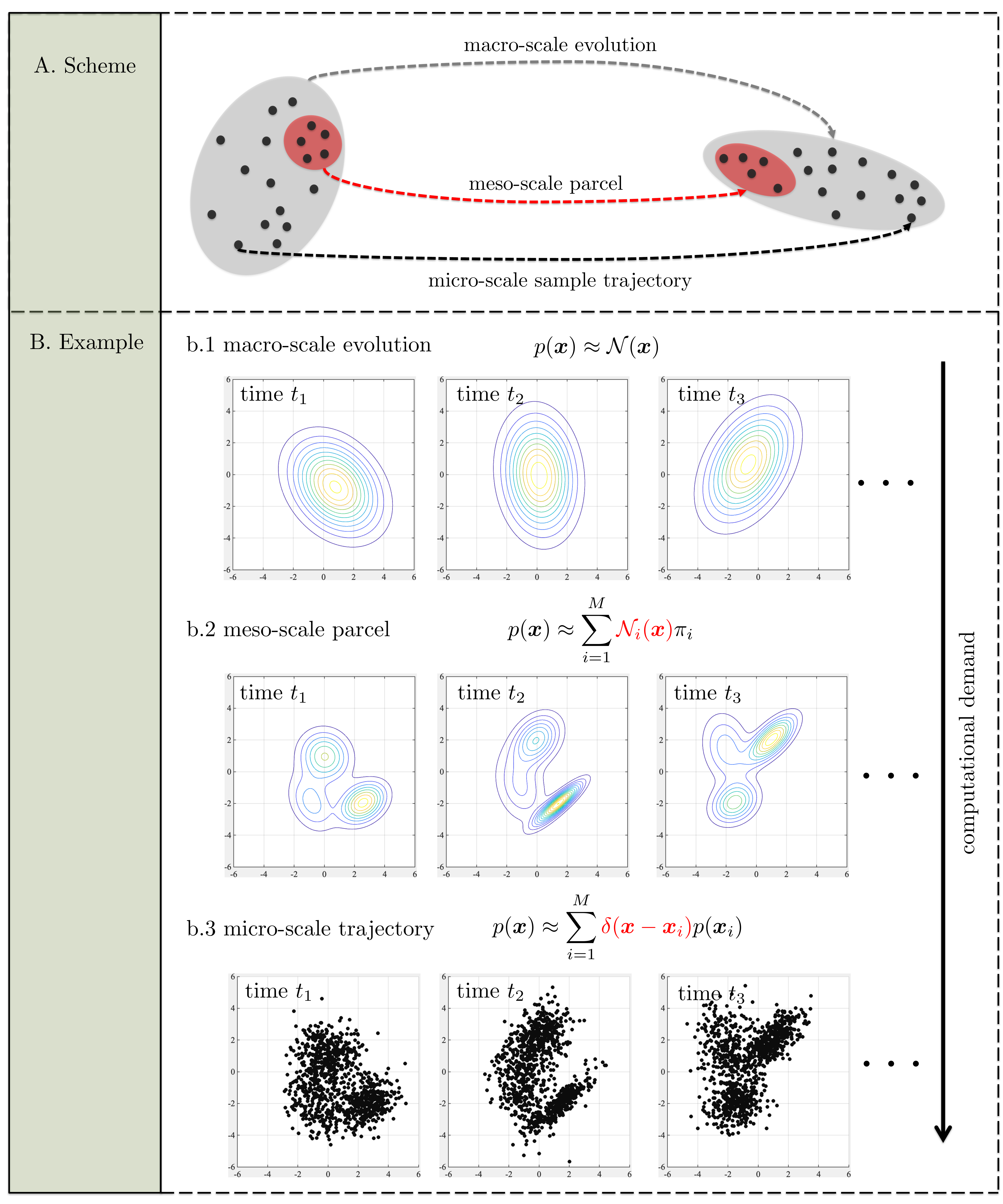}
    \caption{Probability evolution from a macro-, meso-, and micro-scale perspective.}
    \label{fig: f1}
\end{figure}

Compared to the aforementioned macro-scale evolution and micro-scale trajectory, this meso-scale parcel scheme has two advantages. First, mixture modeling can provide asymptotically converged approximations to a given arbitrary probability distribution \cite{bishop2006pattern, kaufman2009finding}. Second, the computational complexity of tracking the evolution of the PDF of interest is much lower than the sampling-based methods where the number of samples required by the MC method in order to adequately estimate the PDF increases significantly \cite{xiu2002wiener, robert2013monte}. Moreover, meso-scale parcel is an effective alternative at an intermediate scale to the macro-/micro-scale methods, where the macro-scale evolution can be regarded as a special case of the meso-scale representation, in which the statistical structure only contains one component, and the micro-scale trajectory can be cast in the meso-scale format with as many components as the number of samples and Dirac delta function as an indicator. \cref{fig: f1} gives a graphic illustration of these three perspectives.

The paper is organized as follows. \cref{sec2} provides the definition of the problems of interest. \cref{sec3} gives the computational guidelines of the proposed meso-scale scheme. \cref{sec4} interprets the meso-scale uncertainty propagation of two special cases: conservative and Markov models. \cref{sec5} provides examples of application to demonstrate the effectiveness of the proposed meso-scale based method. Finally, conclusion is drawn in \cref{sec6} with some discussion on the relative advantages and limitations of the proposed scheme, and potential future works.

\section{Methodology: a meso-scale probability evolution scheme}
\label{sec2}
For notational brevity, let us formulate a stochastic dynamical system in a general model form \cite{strogatz2001nonlinear}:

\begin{equation}
    \label{eq: sec2_1}
    \boldsymbol{x}_t = \mathcal{M} (\boldsymbol{\theta}, t)
\end{equation}

where $\boldsymbol{\theta} = [\theta_1, \theta_2, \dots, \theta_q] \in \mathbb{R}^q$ is an input random vector and $\boldsymbol{x}_t = [x_1, x_2, \dots, x_n] \in \mathbb{R}^n$ represents the corresponding output at time instance $t$, which also is a random vector. 

The goal is to obtain the evolutionary PDF of $\boldsymbol{x}$. Hence, a governing equation with respect to the instantaneous PDF $p(\boldsymbol{x}, t)$ should first be derived \cite{li2004probability}. In the case of continuous random variables, such PDF can be written in a convolution form as \cite{li2009stochastic}:

\begin{equation}
    \label{eq: sec2_2}
    p \left( \boldsymbol{x}_t \right) = (g \star h) (t) = \int_{\Omega_{t}} g(t - \tau) h (\tau) \mathrm{d} \tau
\end{equation}

Solving \cref{eq: sec2_2} in the context of the physical model stated in \cref{eq: sec2_1} using the principle of probability preservation gives the integral expression of the instantaneous PDF in the random input space:

\begin{equation}
    \label{eq: sec2_3}
    p \left( \boldsymbol{x}_t \right) = \int_{\Omega_{\theta}} \delta\left[\boldsymbol{x}_{t} - \mathcal{M}(\boldsymbol{\theta}, t)\right] p(\boldsymbol{\theta}) \mathrm{d} \boldsymbol{\theta}
\end{equation}

where $\delta(\cdot)$ is the Kronecker delta function and $p(\boldsymbol{\theta})$ is the joint PDF describing the input uncertainties, which assumed to be available in this work.

The main point is to find a computationally efficient representation of $p(\boldsymbol{\theta})$ to compute the evolutionary PDF of $\boldsymbol{x}$. This naturally leads to a geometric segmentation, and mixture modeling that learns a probabilistic model by assuming the data are generated from a mixture of distributions is adopted here \cite{bishop2006pattern, kaufman2009finding}. Hence, $p(\boldsymbol{\theta})$ is approximated through a convex combination of a series of weighted base distributions as:

\begin{equation}
    \label{eq: sec2_4}
    p \left( \boldsymbol{\theta} \right) = \sum_{k=1}^{K} \pi_{k} \, \kappa_{k} \left( \boldsymbol{\theta} \right)
\end{equation}

with $\pi_{k}$ denoting the weighting factor for the $k^{th}$ base distribution. Additionally, equality along with box constraints of $\pi_{k}$ are imposed as:

\begin{equation}
    \label{eq: sec2_5}
    \sum_{k=1}^{K} \pi_{k} = 1 \, \, \, \, \, \text{where} \, \, \, \, \, \left( 0 \leqslant \pi_{k} \leqslant 1 \right)
\end{equation}

In practice, normal or multivariate normal distribution function is extensively used as Gaussian mixture model (GMM) is capable of approximating any arbitrary density function when sufficient base terms have been included \cite{dempster1977maximum, mclachlan2007algorithm}. Accordingly, \cref{eq: sec2_4} can be rewritten as:

\begin{equation}
    \label{eq: sec2_6}
    p \left( \boldsymbol{\theta} \right) = \sum_{k=1}^{K} \pi_{k} \, \mathcal{N} \left( \boldsymbol{\theta} | \mu_{k}, \Sigma_{k} \right)
\end{equation}

Henceforth, the rest work is to use optimization methods to identify the parameters $\{ \mu_{k}, \Sigma_{k} \}_{k=1}^{K}$ and substitute optimized parameters to \cref{eq: sec2_3} to compute the instantaneous PDF:

\begin{equation}
    \label{eq: sec2_7}
    p \left( \boldsymbol{x}_t \right) = \sum_{k=1}^{K} \pi_{k} \, \mathcal{K}_{k} \left( \boldsymbol{x}_t \right) \quad \text{where} \quad \mathcal{K}_{k} = \int_{\Omega_{\theta}} \delta\left[\boldsymbol{x}_{t} - \mathcal{M}(\boldsymbol{\theta}, t)\right] \mathcal{N}_{k} \left( \boldsymbol{\theta}\right)\mathbf{d} \boldsymbol{\theta}
\end{equation}

\section{Computational algorithms and implementation details}
\label{sec3}
The computational algorithm of the proposed meso-scale scheme contains two major steps. First, one has to determine the weight, location, and scale of each meso-scale component. Second, the optimized meso-scale components are integrated into a single kernel by \cref{eq: sec2_7}. To ensure the precision of the constructed kernel regarding describing the dynamical behaviors, the number of the component PDFs has to be sufficiently large \cite{kurtz2013cross, kaufman2009finding}. As a result, effective optimization of model parameters is a prerequisite in this case as conventional algorithms such as expectation-maximization by definition involve the optimization of a larger number of parameters \cite{dempster1977maximum, mclachlan2007algorithm}.

To reduce the computational complexity, weighting factors are assumed to be homogeneous, that is, $K$ in \cref{eq: sec2_4} is assumed sufficiently large, and hence we have $\widehat{\pi}_{k}=\frac{1}{K}(k=1, \dots, K)$. Next, the locations $\{ \mu_{k} \}_{k=1}^{K}$ can be directly determined by either a low-discrepancy sequence or clustering algorithms (\cref{sec31}), and the only parameters that needed to be optimized at this stage are $\{ \sigma_{k} \}_{k=1}^{K}$ and evaluation of this reduced model can be efficiently completed by any standard optimization algorithm (\cref{sec32}). Moreover, spherical-radial integration is introduced to compute the evolutionary kernel, where a third-degree spherical-radial cubature rule is adopted for efficient numerical integration (\cref{sec33}).

\subsection{Determination of the representative points}
\label{sec31}
\begin{figure}[h!]
    \centering
    \includegraphics[width=1.0\textwidth]{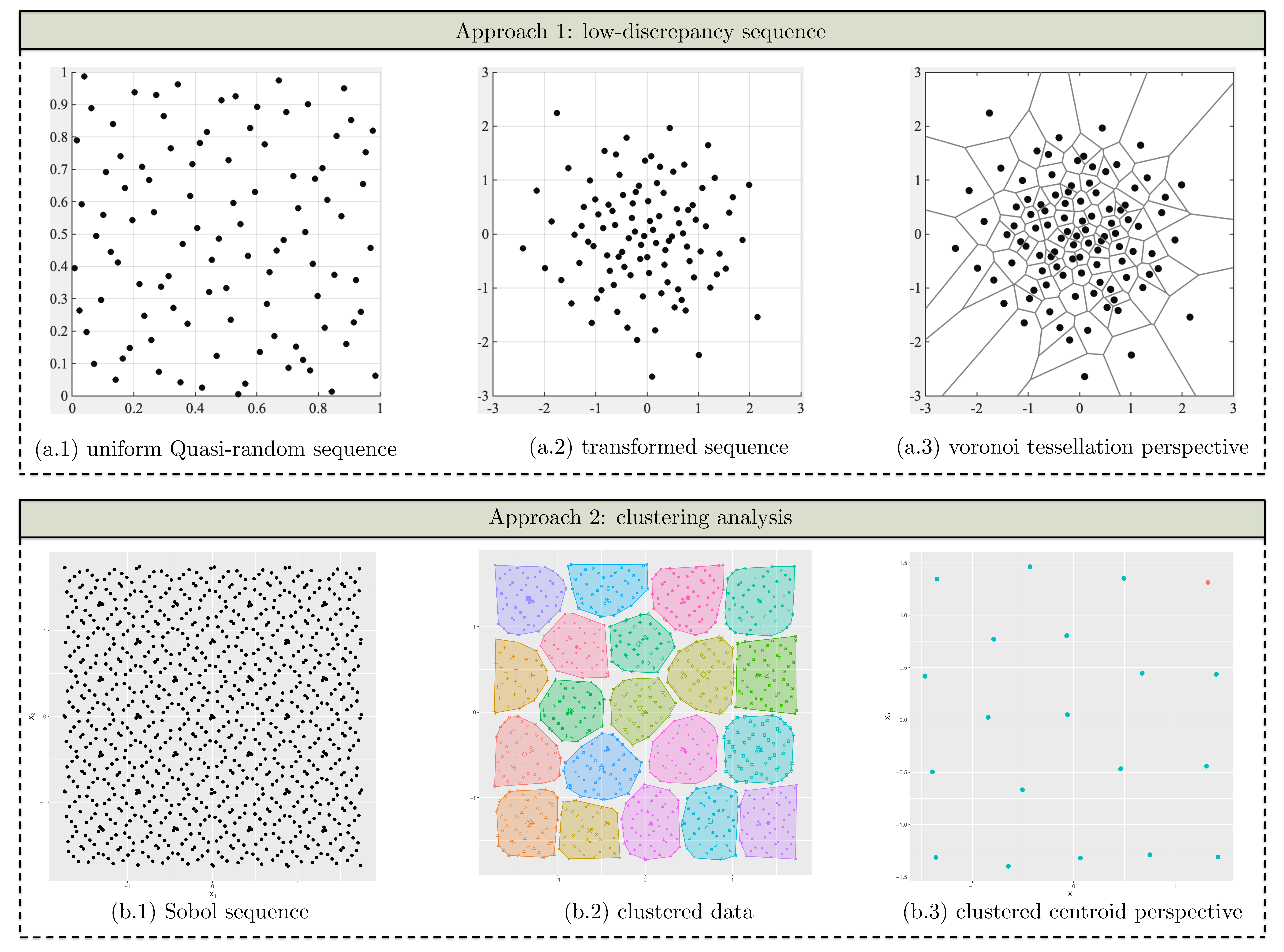}
    \caption{Illustration of using low discrepancy sequences and clustering methods to compute representative points.}
    \label{fig: f2}
\end{figure}

To illustrate the connections between location parameters and tessellation/cluster centroids, let us consider an open set $\Omega_{\theta} \subseteq \mathbb{R}^{q}$. The sub-domain set $\{ V_k \}_{k=1}^{K}$ is called a tessellation of $\Omega_{\theta}$ if $V_i \cap V_j = \varnothing$ for $i \neq j$ and $\cup_{k=1}^{K} V_k = \Omega_{\theta}$. Each is $V_k$ assigned to a point $\theta_k$. The representative point set (or rep-point set for brevity) $\mathcal{P} = \{ \theta_k \}_{k=1}^{K}$ corresponds to a tessellation of $\Omega_{\theta}$ and also provides a candidate for $\{ \mu_{\theta_k} \}_{k=1}^{K}$. The task of finding $\mathcal{P}$ can be approximately performed by finding the best partition of $\Omega_{\theta}$ \cite{bishop2006pattern, kaufman2009finding}. A measure of the quality of the partition can be given by the quadrature error that is defined as \cite{du1999centroidal}:

\begin{equation}
    \label{eq: sec3_1}
    \begin{aligned}
        \mathcal{E}(\boldsymbol{\theta}) & =\left\|\int_{\Omega_{\theta}} p(\boldsymbol{\theta}) \mathrm{d} \boldsymbol{\theta}-\sum_{k=1}^{N} \pi_{k} p\left(\boldsymbol{\theta}_{k}\right) \right\| \leq L \sum_{k=1}^{K} \int_{V_k}\left\|\boldsymbol{\theta}-\boldsymbol{\theta}_k\right\| \mathrm{d} \boldsymbol{\theta}
    \end{aligned}
\end{equation}

where $|| \cdot ||$ denotes a general norm and $L$ is a Lipschitz constant. If measured through information discrepancy using entropy theory, the quality of partition stated in \cref{eq: sec3_1} can be reformulated as:

\begin{equation}
    \label{eq: sec3_2}
    \mathcal{D}_{\mathcal{F}} (\mathcal{P}) = \sup_{\theta \in \Omega_{\theta}} \left| \frac{1}{K} \sum_{k=1}^{K} \mathbb{I} \{ \theta_k \leq \theta \} - \mathcal{F}(\theta) \right|
\end{equation}

where $\mathcal{F} (\boldsymbol{\theta})$ is the cumulative distribution function (CDF) of $\boldsymbol{\theta}$ and $\mathbb{I}(\cdot)$ is the indicator function. This $\mathcal{F}$-discrepancy coincides with the Kolmogorov-Smirnov distance \cite{illian2008statistical}. Hence, $\mathcal{P}$ can be determined in two ways: through a low discrepancy sequence by minimizing the $\mathcal{F}$-discrepancy (\cref{sec311}), or through clustering points by minimizing the quadrature error (\cref{sec312}). \cref{fig: f2} provides a graphic illustration of these two approaches.

\subsubsection{Low-discrepancy sequence approach}
\label{sec311}
The concept of low discrepancy sequence (LDS) is central to the quasi-Monte Carlo  method (QMC) and to the number theoretic method (NTM) \cite{robert2013monte, zhang2013structural, illian2008statistical}. An LDS is basically a uniformly distributed point set $\mathcal{U} = \{ \boldsymbol{\mu}_k \}_{i=k}^{k}$ generated in a unit hypercube $\mathcal{C}^q = [0, 1]^q$, where $q$ is the dimension of the input random space. The rep-point set $\mathcal{P}$ for distributions different from the uniform can be obtained via transformation as:

\begin{equation}
    \label{eq: sec3_3}
    \mathcal{P} = T(\mathcal{U})
\end{equation}

where $T(\cdot)$ is a transformation function. Several effective LDSs, e.g. the good lattice point set (GLP), good point set (GP), Halton sequence, Haber sequence, Hammersley sequence, Faure sequence, Sobol sequence, etc, have been shown to provide lower $\mathcal{F}$-discrepancies than random point sets \cite{nie2004new, sudret2008global, wei2015variable, zhang2013structural}. $T(\cdot)$ may be simply taken as the inverse of the CDF of $\boldsymbol{\theta}$, that is $T(\cdot) = \mathcal{F}^{-1}(\cdot)$. Generally, $\mathcal{F}^{-1}(\boldsymbol{\theta})$ has independent marginal, i.e. $\mathcal{F}^{-1}(\boldsymbol{\theta}) = \mathcal{F}^{-1}(\theta_1, \theta_2, \dots, \theta_q) = \Pi_{i=1}^{q} \mathcal{F}^{-1}(\theta_i)$. Therefore,

\begin{equation}
    \label{eq: sec3_4}
    \boldsymbol{\theta}_i = [ \mathcal{F}^{-1}(\mu_1), \mathcal{F}^{-1}(\mu_2), \dots, \mathcal{F}^{-1}(\mu_q) ]
\end{equation}

where the $\mathcal{F}$-discrepancy of $\mathcal{P}$ concerning $\mathcal{F} (\boldsymbol{\theta})$ is the same as the one for $\mathcal{F} (\boldsymbol{\mu})$. Techniques for generating $\mathcal{P}$ for elliptically contoured and multivariate Liouville distributions can be found in the literature \cite{fang2018symmetric}. In particular, for a multivariate standard Gaussian distribution, the Box-Muller transformation provides a good alternative to the method based on the inverse of the CDF. Then, $\mathcal{P}$ can be taken as the means of the components of a GMM, i.e. $\{ \widehat{\boldsymbol{\mu}_{\theta_K}} \}_{K=1}^{K} = \mathcal{P}$. An advantage of this manipulation is that the rep-point set transformed from an LDS can make the components distributed more uniformly than a random point set. A GLP set on $\mathcal{C}^2$ is shown in \cref{fig: f2}. (a.1), its transformation to a Gaussian distribution is shown in \cref{fig: f2}. (a.2), and the corresponding voronoi tessellation is provided in \cref{fig: f2}. (a.3).

\subsubsection{Clustering analysis approach}
\label{sec312}
While the transformation of an LDS usually suits independent random variables, clustering methods provide an approach for dependent random variables \cite{kaufman2009finding}. In essence, clustering methods aim to minimize the last term of \cref{eq: sec3_1} instead of minimizing the $\mathcal{F}$-discrepancy \cite{dempster1977maximum, mclachlan2007algorithm}. For example, for the K-means clustering algorithm, this minimization is equivalent to seeking the mass centroid of $V_k$, i.e.

\begin{equation}
    \label{eq: sec3_5}
    \boldsymbol{\theta}_{k} = \frac{\int_{V_{k}} \boldsymbol{\theta p}(\boldsymbol{\theta}) \mathrm{d} \boldsymbol{\theta}}{\int_{V_{k}} p(\boldsymbol{\theta}) \mathrm{d} \boldsymbol{\theta}}
\end{equation}

and it can be implemented as follows: (1) Initialize the cluster centers $\mathcal{P} = \{ \boldsymbol{\theta}_{i} \}_{i=1}^{N}$ from $p(\boldsymbol{\theta})$ by randomly sampling from $p(\boldsymbol{\theta})$; (2) Randomly sample an auxiliary point set $\{ \boldsymbol{\theta}_{j}^{'} \}_{j=1}^{M}$ from $p(\boldsymbol{\theta})$, where $M \gg N$; (3) Assign $\boldsymbol{\theta}_{1}^{'}, \boldsymbol{\theta}_{2}^{'}, \dots, \boldsymbol{\theta}_{M}^{'}$ respectively to their nearest cluster centers based on the Eulerian distance; (4) Update each center $\boldsymbol{\theta}_{i}$ with the mean of the auxiliary points assigned to it; (5) Repeat step (3) and (4) until $\mathcal{P}$ no longer changes. The second row of \cref{fig: f2} graphically illustrates the connection between cluster centroids and the proposed meso-scale statistical structures.

\subsection{Determination of the Gaussian components}
\label{sec32}
\begin{figure}[h!]
    \centering
    \includegraphics[width=1.0\textwidth]{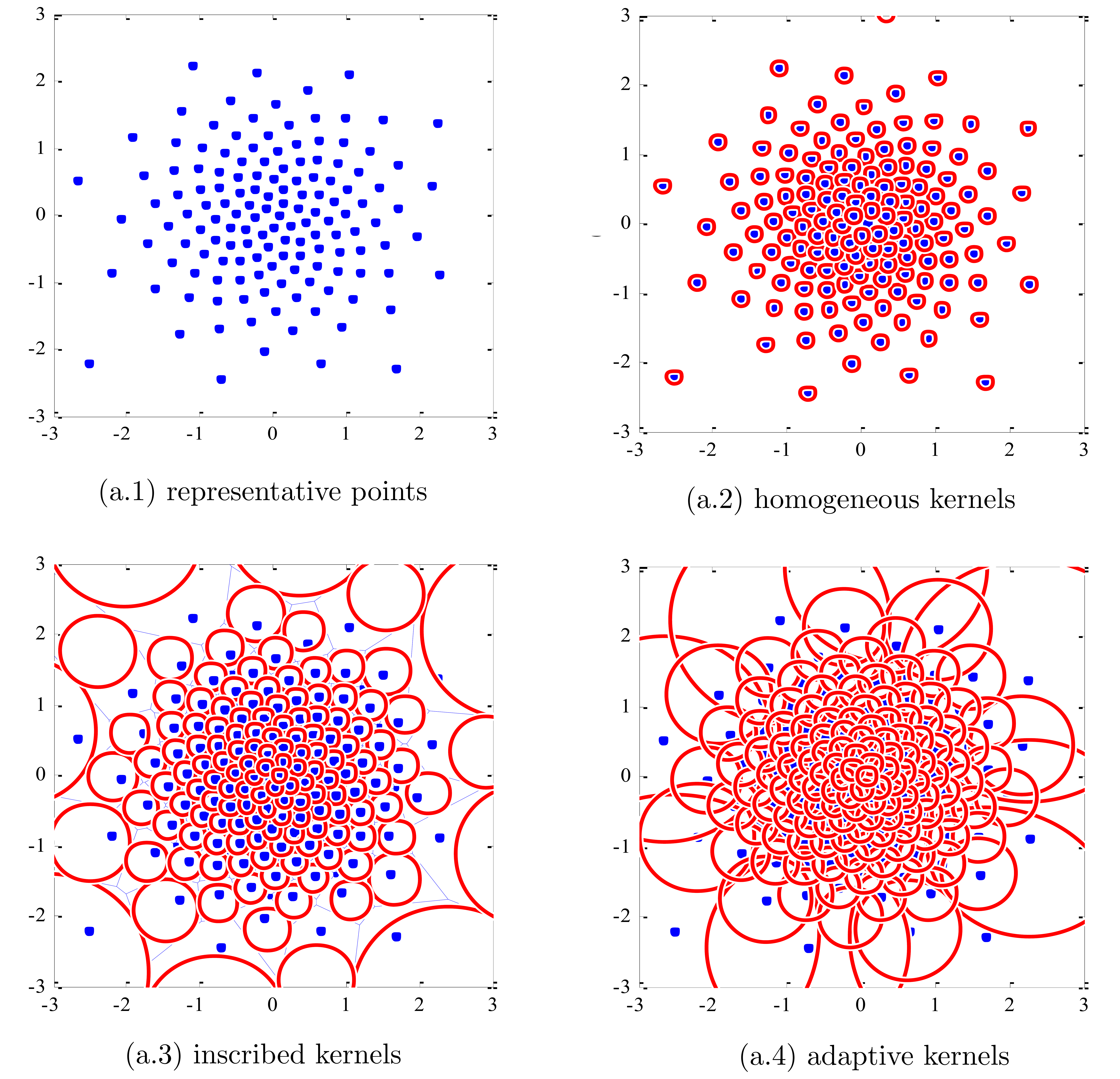}
    \caption{Illustration of the homogeneous kernel, inscribed kernel, and adaptive kernel.}
    \label{fig: f3}
\end{figure}

With computed location parameters, the approximation capability of the constructed mixture model is primarily determined by the complexity of the selected kernel \cite{bishop2006pattern, kaufman2009finding}. There are three types of kernels, namely, homogeneous kernel, inscribed kernel, and adaptive kernel. \cref{fig: f3} gives a summary of these kernels.

To demonstrate the influences that model configuration has on the approximation capability, we resort to the concept of Gaussian complexity, which is defined as \cite{eldan2018gaussian}:

\begin{equation}
    \label{eq: sec3_6}
    \mathcal{G} (\mathcal{T}) :=\mathbb{E} \sup_{t \in \mathcal{T}} \frac{1}{n} \sum_{i=1}^{n} \xi_{i} t_{i}
\end{equation}

where $\xi_{1}, \xi_{2}, \dots, \xi_{n}$ are independent standard Gaussian random variables and $\mathcal{T} \subseteq \mathbb{R}^n$ represents the set of interest. 

\begin{figure}[h!]
    \centering
    \includegraphics[width=1.0\textwidth]{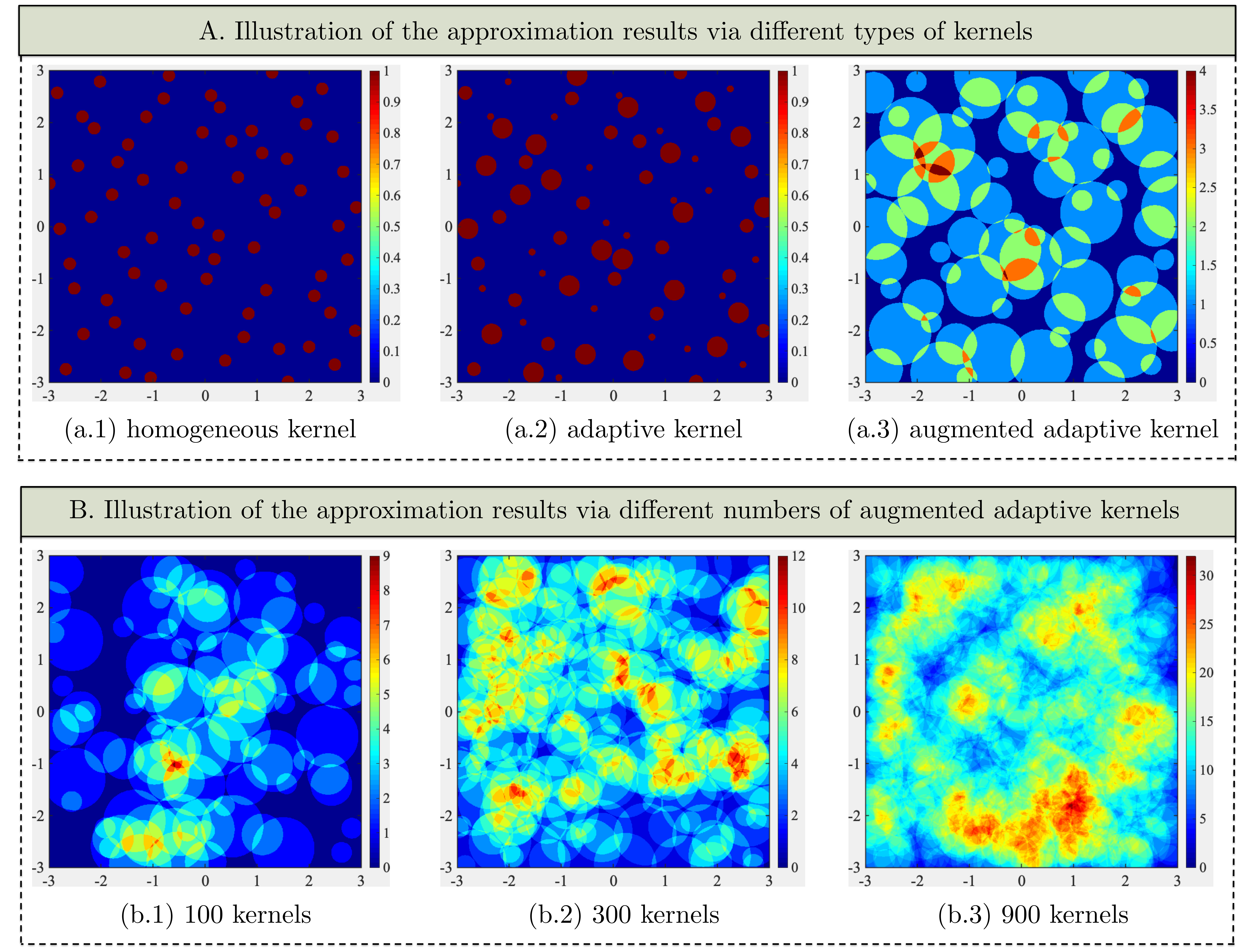}
    \caption{Illustration of the approximation results via different kernels and different number of kernels.}
    \label{fig: f4}
\end{figure}

According to \cite{eldan2018gaussian}, the approximation space of interest to us is a function $f(\cdot)$ takes values of order $O(n)$ and whose Lipschitz constant is $\text{Lip} = O(1)$. It is clear that such functions trivially have complexity at most $O(n)$. Hence, the kernel type and kernel numbers are the most important factors that should be addressed in the implementation. \cref{fig: f4} graphically illustrates the influences of these two factors. Note that each circle represents a Gaussian component, for instance, homogeneous kernels indicate the same variance has been assigned. Also, the heatmap comparison is based on the complexity of the configured model using \cref{eq: sec3_6}.

For numerical implementation, the optimal set of $\{ \boldsymbol{\mu}_k \}_{i=k}^{k}$ is assumed given by $\mathcal{P}$. The optimal set of $\{ \boldsymbol{\Sigma}_k \}_{i=k}^{k}$ can be determined by the EM algorithm. It should be observed that EM could be used to determine all the parameters, including the weights $\{ \omega_k \}_{i=k}^{k}$, means $\{ \boldsymbol{\mu}_k \}_{i=k}^{k}$ and covariances $\{ \boldsymbol{\Sigma}_k \}_{i=k}^{k}$ \cite{dempster1977maximum, mclachlan2007algorithm}. However, as anticipated, in this case $\{\omega_k \}_{i=k}^{k}$ and $\{ \boldsymbol{\mu}_k \}_{i=k}^{k}$ are determined in advance, therefore significantly reducing the complexity of the optimization problem and allowing the use of more kernels. The corresponding EM algorithm can be implemented according to the steps stated in \cref{app1}.

\subsection{Determination of the evolutionary kernels}
\label{sec33}
Similarly to what is done in macro-scale methods, the second-order statistics can be used to describe each evolutionary kernel given by \cref{eq: sec2_7}. Specifically, the objective is the estimation of the first two statistical moments of the system states $\boldsymbol{x}_t$, i.e., the mean

\begin{equation}
    \label{eq: sec3_7}
    \mu_k (\boldsymbol{x}_t) = \int_{\Omega_\theta} \mathcal{M}(\boldsymbol{\theta}, t) \mathcal{N}_{k} \left( \boldsymbol{\theta} \right)\mathbf{d} \boldsymbol{\theta} \approx \sum_{i=1}^{N} \pi_{i,k} \mathcal{M}(\boldsymbol{\theta}_{i,k}, t)
\end{equation}

and the variance

\begin{equation}
    \label{eq: sec3_8}
    \begin{aligned}
    \sigma_k (\boldsymbol{x}_t) & = \int_{\Omega_\theta} \left[ \boldsymbol{x}_t - \mu_k (\boldsymbol{x}_t) \right] \left[ \boldsymbol{x}_t - \mu_k (\boldsymbol{x}_t) \right]^{T} \mathcal{N}_{k} \left( \boldsymbol{\theta} \right) \mathbf{d} \boldsymbol{\theta} \\
    & \approx \sum_{i=1}^{N} \pi_{i,k} \left[ \mathcal{M}(\boldsymbol{\theta}_{i,k}, t) - \mu_k (\boldsymbol{x}_t) \right] \left[ \mathcal{M}(\boldsymbol{\theta}_{i,k}, t) - \mu_k (\boldsymbol{x}_t) \right]^{T}
    \end{aligned}
\end{equation}

where $N$ denotes the number of auxiliary points assigned to the $k^{th}$ evolutionary kernel and $\boldsymbol{\theta}_{i,k}$ denotes the $j^{th}$ auxiliary point with weight $\pi_{i,k}$ for $\mathcal{K}_k(\boldsymbol{x}_t)$.

The auxiliary points can be chosen from an LDS, a random point set, a sigma-point set or a cubature point set \cite{illian2008statistical}. A Gaussian cubature point set is favorable here since the integrand $\mathcal{N}_{k} \left( \boldsymbol{\theta} \right)$ is a Gaussian density function. A Gaussian density function has a symmetric and radial shape. This feature facilitates efficient integration rules \cite{arasaratnam2009cubature, jia2013high}. Explicitly, an integral with the integrand of the form

\begin{equation}
    \label{eq: sec3_9}
    I(\mathbf{f})=\int_{\mathcal{D}} \mathbf{f}(\mathbf{x}) w(\mathbf{x}) d \mathbf{x}
\end{equation}

where $\mathbf{f}(\cdot)$ is some arbitrary nonlinear function, $\mathcal{D} \subseteq \mathbb{R}^{n}$ denotes the region of integration, and $w(\mathbf{x})$ is the known weighting function can be transformed into a spherical-radial integration. Therefore, \cref{eq: sec3_7} can be expressed as:

\begin{equation}
    \label{eq: sec3_10}
    \int_{\Omega_{\theta}} \mathcal{M}(\boldsymbol{\theta}) e^{-\boldsymbol{\theta}^{T} \boldsymbol{\theta}} \mathrm{d} \boldsymbol{\theta} \approx \sum_{p=1}^{M_{s}} \sum_{q=1}^{M_{r}} a_{q} b_{p} \mathcal{M}\left(r_{q} s_{p}\right)
\end{equation}

with

\begin{equation}
    \label{eq: sec3_11}
    \boldsymbol{\theta} = \boldsymbol{r} \boldsymbol{s} \quad \textit{and} \quad \boldsymbol{s}^T \boldsymbol{s} = 1
\end{equation}

where $M_{s} \times M_{r}$ represents the number of cubature points used in the spherical-radial cubature, while $a_{q}$ and $b_{p}$ are constants to be determined \cite{arasaratnam2009cubature, jia2013high}. If $\mathcal{N} \left( \boldsymbol{\theta} \right)$ has mean $\mu_{\boldsymbol{\theta}}$ and covariance $\Sigma_{\boldsymbol{\theta}}$, the above integral can be further simplified to:

\begin{equation}
    \label{eq: sec3_12}
    \int_{\Omega_\theta} \mathcal{M}(\boldsymbol{\theta}) \mathcal{N} \left( \boldsymbol{\theta} \right)\mathbf{d} \boldsymbol{\theta} = \frac{1}{\sqrt{\pi^q}} \int_{\Omega_\theta} \mathcal{M}( \sqrt{2 \Sigma_{\boldsymbol{\theta}}}\boldsymbol{\theta} + \mu_{\boldsymbol{\theta}}) e^{-\boldsymbol{\theta}^{T} \boldsymbol{\theta}} \mathrm{d} \boldsymbol{\theta}
\end{equation}

Through a third-degree spherical-radial cubature rule with $\mu_{\boldsymbol{\theta}} = \boldsymbol{0}$ and covariance $\Sigma_{\boldsymbol{\theta}} = \boldsymbol{I}$, \cref{eq: sec3_12} writes as:

\begin{equation}
    \label{eq: sec3_13}
    \int_{\Omega_\theta} \mathcal{M}(\boldsymbol{\theta}) \mathcal{N} \left( \boldsymbol{\theta} \right)\mathbf{d} \boldsymbol{\theta} = \sum_{j=1}^{2q} \gamma_j \mathcal{M}(\xi_j)
\end{equation}

with

\begin{equation}
    \label{eq: sec3_14}
    M_{r} = 1 \quad \text{and} \quad M_{s} = 2q \quad \text{and} \quad \gamma_j = \frac{1}{2q} \quad \text{and} \quad \xi_j = \sqrt{q} \llcorner 1 \lrcorner
\end{equation}

Note $j = 1, 2, \dots, 2q$ where $q$ represents the dimension of the input random space and $\llcorner 1 \lrcorner$ denotes the
vector:

\begin{equation}
    \label{eq: sec3_15}
    \llcorner 1 \lrcorner = \left[ 0, 0, \dots, h_j, \dots, 0, 0 \right]
\end{equation}

where $j^{th}$ component $h=1$ or $h=-1$. A comparison of \cref{eq: sec3_13} with \cref{eq: sec3_7} and \cref{eq: sec3_8} yields:

\begin{equation}
    \label{eq: sec3_16}
    N = 2q \quad \text{and} \quad \pi_{i,k} = \frac{1}{2q} \quad \text{and} \quad \boldsymbol{\theta}_{i,k} = \sqrt{\Sigma_{\boldsymbol{\theta}_k}}\xi_i+\mu_{\boldsymbol{\theta}_k}
\end{equation}

\begin{figure}[h!]
    \centering
    \includegraphics[width=1.0\textwidth]{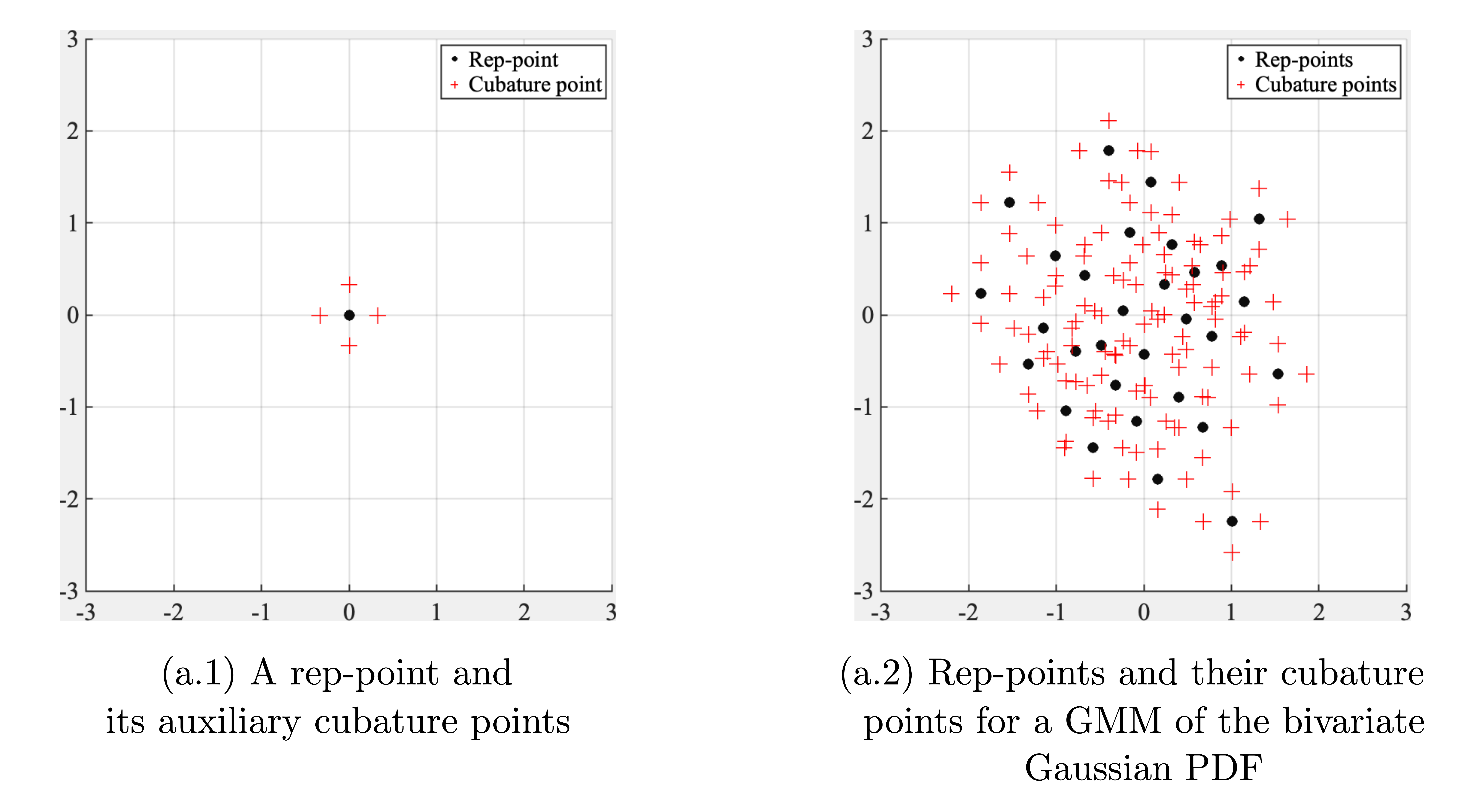}
    \caption{Illustration of the rep-points and the auxiliary cubature points.}
    \label{fig: f5}
\end{figure}

Substituting these back into \cref{eq: sec3_7} and \cref{eq: sec3_8} allows the determination of the $k^{th}$ evolutionary kernel. The global probability evolution is obtained by assembling all the evolutionary  kernels. \cref{fig: f5}. (a) illustrates a 2-dimensional cubature point set of a component PDF. The rep-points and their corresponding cubature points of a GMM of the bivariate Gaussian PDF are shown in \cref{fig: f5}. (b).

\section{Meso-scale perspective of two classical typologies of probability evolution}
\label{sec4}

\subsection{Meso-scale scheme for conservative models}
\label{sec41}
A conservative model places the randomness of a stochastic system in an augmented random initial condition. Particularly, random model parameters $\boldsymbol{Y}(t) = [Y_1(t), Y_2(t), \dots, Y_m(t)]$ can be written in a state space where $\boldsymbol{Y}(t)$ is derived from a random initial condition $\boldsymbol{Y}(t_0)$ via deterministic dynamics \cite{li2009stochastic, ghanem2003stochastic}. The random excitations $\boldsymbol{W}(t) = [W_1(t), W_2(t), \dots, W_v(t)]$ can also be written in a state space by a series expansion, where the random coefficients are regarded as random initial conditions and the deterministic basis functions are regarded as deterministic dynamics \cite{melchers2018structural, xiu2002wiener}. Thus, $\boldsymbol{Y}(t)$, $\boldsymbol{W}(t)$, and $\boldsymbol{X}(t)$ can be assembled and formulated in an augmented state as:

\begin{equation}
    \label{eq: sec4_1}
    \boldsymbol{Z}(t) = [\boldsymbol{Y}(t), \boldsymbol{W}(t), \boldsymbol{X}(t)]^{T}
\end{equation}

with random initial condition

\begin{equation}
    \label{eq: sec4_2}
    \boldsymbol{Z}(t_0) = \boldsymbol{Z}_0 = [\boldsymbol{Y}_0, \boldsymbol{W}_0, \boldsymbol{X}_0]^{T}
\end{equation}

It is understood that a dynamic stochastic system with given states (\cref{eq: sec4_1}) and initial conditions (\cref{eq: sec4_2}) can be expressed in a Lagrangian differential format \cite{grigoriu2013stochastic}:

\begin{equation}
    \label{eq: sec4_3}
\left\{\begin{array}{l}\dot{\boldsymbol{Z}}(t)=\boldsymbol{g}_{\mathrm{aug}}\left(\boldsymbol{Z}_{0}, t\right) \\ \boldsymbol{Z}(t)=\boldsymbol{h}_{\mathrm{aug}}\left(\boldsymbol{Z}_{0}, t\right)\end{array} ; \boldsymbol{Z}\left(t_{0}\right)=\boldsymbol{Z}_{0}=\left[\begin{array}{l}\boldsymbol{X}_{0} \\ \boldsymbol{Y}_{0}\end{array}\right]\right.
\end{equation}

where $\boldsymbol{g}_{\mathrm{aug}}$ and $\boldsymbol{h}_{\mathrm{aug}}$ are respectively the augmented velocity function and state function of initial condition $\boldsymbol{Z}_{0}$.

To compute the probability evolution of a conservative model from a meso-scale perspective (\cref{eq: sec2_4}), the input PDF is expressed as:

\begin{equation}
    \label{eq: sec4_4}
    p\left(\mathbf{Z}_{t_{0}}\right)=\sum_{i=1}^{N} \kappa_{\mathrm{aug}}^{(i)}\left(\mathbf{Z}_{t_{0}}\right) \omega_{\mathrm{aug}}^{(i)}
\end{equation}

Correspondingly, the evolutionary PDF is derived in a similar manner to \cref{eq: sec2_7} and it writes as:

\begin{equation}
    \label{eq: sec4_5}
    p\left(\mathbf{Z}_{t_{j}}\right) = \sum_{i=1}^{N} \omega_{\mathrm{aug}}^{(i)} \mathcal{K}_{\mathrm{aug}}^{(i)}\left( \mathbf{Z}_{t_{j}} \right)
\end{equation}

where

\begin{equation}
    \label{eq: sec4_6}
    \mathcal{K}_{\mathrm{aug}}^{(i)}\left( \mathbf{Z}_{t_{j}} \right) = \int_{\Omega_{\mathrm{aug}, t_{0}}} p\left(\mathbf{Z}_{t_{j}} | \mathbf{Z}_{t_{0}}\right) \kappa_{\mathrm{aug}}^{(i)}\left(\mathbf{Z}_{t_{0}}\right) \mathrm{d} \mathbf{Z}_{t_{0}}
\end{equation}

Using the third-order Gaussian cubature rule \cite{arasaratnam2009cubature}, the $i^{th}$ evolutionary kernel is represented with mean

\begin{equation}
    \label{eq: sec4_7}
    \mu^{(i)}\left(\boldsymbol{Z}_{t_{j}}\right) \approx \frac{1}{N^{(i)}} \sum_{k=1}^{N^{(i)}} \boldsymbol{h}_{\mathrm{aug}}\left(\mathbf{Z}_{t_{0}}^{(i, k)}, t_{j}-t_{0}\right)
\end{equation}

and covariance

\begin{equation}
    \label{eq: sec4_8}
    \Sigma^{(i)}\left(\boldsymbol{Z}_{t_{j}}\right) \approx \frac{1}{N^{(i)}} \sum_{k=1}^{N^{(i)}} L_{\mathrm{aug}, t_{j}}^{(i, k)} {L_{\mathrm{aug}, t_{j}}^{(i, k)}}^{T}
\end{equation}

where

\begin{equation}
    \label{eq: sec4_9}
    L_{\mathrm{aug}, t_{j}}^{(i, k)}= \boldsymbol{h}_{\mathrm{aug}}\left(\mathbf{Z}_{t_{0}}^{(i, k)}, t_{j}-t_{0}\right)-\mu^{(i)}\left(\mathbf{Z}_{t_{j}}\right)
\end{equation}

\begin{remark}[1]
    In the view of state-space modeling, meso-scale scheme deals with conservative models as a nonlinear filter, i.e. cubature Kalman filter (CKF). The multi-dimensional integrals involved in the time(predictive density) and measurement(posterior density) Bayesian updating of the CKF are efficiently addressed via the cubature rule. Such a derivative-free method broadens the applicability of the proposed meso-scale scheme \cite{wan2000unscented, arasaratnam2009cubature, jia2013high}.
\end{remark}

\subsection{Meso-scale scheme for Markov models}
\label{sec42}
The second type of model that is closely connected to the proposed meso-scale scheme is the Markov model \cite{grigoriu2013stochastic}. By definition, Markov models posit that the randomness of a stochastic system is injected sequentially by random excitations. The initial condition may also be random. Usually, a Markov model is written as

\begin{equation}
    \label{eq: sec4_10}
\left\{\begin{aligned} \mathrm{d} \boldsymbol{X}_{t} &=G\left(\boldsymbol{X}_{t}, t\right) \mathrm{d} t+A\left(\boldsymbol{X}_{t}, t\right) \mathrm{d} \boldsymbol{B}_{t} \\ \boldsymbol{X}_{t}=& \boldsymbol{X}_{0}+\int_{t_{0}}^{t} G\left(\boldsymbol{X}_{s}, s\right) \mathrm{d} s+\int_{t_{0}}^{t} A\left(\boldsymbol{X}_{s}, s\right) \mathrm{d} \boldsymbol{B}_{s} \end{aligned}\right.
\end{equation}

where $\mathrm{d} \boldsymbol{B}_{t}$ dentoes the random excitation. It is a Wiener process with mean $\mu(\mathrm{d} \boldsymbol{B}_{t})=0$ and covariance $\Sigma(\mathrm{d} \boldsymbol{B}_{t})=\boldsymbol{D} \mathrm{d} t$. Moreover, such a Wiener process holds:

\begin{equation}
    \label{eq: sec4_11}
    \boldsymbol{W}_t = \frac{\mathrm{d} \boldsymbol{B}_{t}}{\mathrm{d} t}
\end{equation}

In a similar manner(\cref{eq: sec2_4}), the input PDF is $p\left(\boldsymbol{X}_{t_{j}-1}\right)$ and it is expressed by the meso-scale representation:

\begin{equation}
    \label{eq: sec4_12}
    p\left(\boldsymbol{X}_{t_{j}-1}\right)=\sum_{i=1}^{N} \omega^{(i)} \mathcal{K}^{(i)}\left(\boldsymbol{X}_{t_{j}}\right) 
\end{equation}

Accordingly, the probability evolution is:

\begin{equation}
    \label{eq: sec4_13}
    \mathcal{K}^{(i)}\left(\boldsymbol{X}_{t_{j}}\right) = \int_{\Omega_{t_{j-1}}} p\left(X_{t_{j}} | X_{t_{j-1}}\right) \kappa^{(i)}\left(X_{t_{j-1}}\right) d X_{t_{j-1}}
\end{equation}

Combine \cref{eq: sec2_7} and \cref{eq: sec4_10}, \cref{eq: sec4_13} can be further expressed as:

\begin{equation}
    \label{eq: sec4_14}
    \begin{aligned}
    \mathcal{K}^{(i)}\left(\boldsymbol{X}_{t_{j}}\right) & = \int_{\Omega_{t_{j-1}}} \Bigg\{\int_{\Omega_{t_{j-1}}} \delta\left[\boldsymbol{X}_{t_{j}}-\boldsymbol{X}_{t_{j}}^{(i)}-\int_{t_{j-1}}^{t_{j}} G\left(\boldsymbol{X}_{s}, s\right) \mathrm{d} s-\int_{t_{j-1}}^{t_{j}} A\left(\boldsymbol{X}_{s}, s\right) \mathrm{d} \boldsymbol{B}_{S}^{(i)}\right] \\
    & p\left(\boldsymbol{w}_{s}\right) \mathrm{d} \boldsymbol{w}_{s} \Bigg\} \times p(X_{t_{j-1}}) d X_{t_{j-1}}
    \end{aligned}
\end{equation}

where $\Omega_{t_{j-1} : t_j}$ denotes the integral domain, including infinite time slices from $t_{j-1}$ to $t_j$. With the derivative-free discretization, the second-order statistics become:

\begin{equation}
    \label{eq: sec4_15}
    \mu^{(i)}\left(X_{t_{j}}\right) \approx \frac{1}{N^{(i)}} \sum_{k=1}^{N_{j}^{(i)}}\left[X_{t_{j-1}}^{(i, k)}+\int_{t_{j-1}}^{t_{j}} G\left(X_{s}, s\right) \mathrm{d} s+\int_{t_{j-1}}^{t_{j}} A\left(X_{s}, s\right) \mathrm{d} B_{s}^{(i, k)}\right]
\end{equation}

and 

\begin{equation}
    \label{eq: sec4_16}
    \Sigma^{(i)}\left(X_{t_{j}}\right) \approx \frac{1}{N^{(i)}} \sum_{k=1}^{N_{j}^{(i)}} L_{t_{j}}^{(i, k)} {L_{t_{j}}^{(i, k)}}^T
\end{equation}

where

\begin{equation}
    \label{eq: sec4_17}
    L_{t_{j}}^{(i, k)}=\left[X_{t_{j-1}}^{(i, k)}+\int_{t_{j-1}}^{t_{j}} G\left(X_{s}, s\right) \mathrm{d} s+\int_{t_{j-1}}^{t_{j}} A\left(X_{s}, s\right) \mathrm{d} B_{s}^{(i, k)}-\mu^{(i)}\left(X_{t_{j}}\right)\right]
\end{equation}

\begin{remark}[2]
    Note that the integral domain is not $\Omega_{t_{j-1}}$ but $\Omega_{t_{j-1}} \times \Omega_{t_{j-1}:t_j}$, which is time-varying. Therefore, a new GMM should be constructed for $p(\boldsymbol{X}_{t_{j-1}}, \boldsymbol{\omega}_{t_{j-1}:s})$ as each random excitation $\omega_s$ is injected \cite{eldan2018gaussian}. Considering that $\boldsymbol{W}_{t_{j-1}:t_j}$ is an additive noise and independent of $\boldsymbol{X}_{t_{j-1}}$, a GMM can be initially constructed for $p(\boldsymbol{X}_{t_{j-1}})$ and then updated according to $\boldsymbol{W}_{t_{j-1}:t_j}$. However, instead of constructing a series of new GMMs between time $t_{j-1}$ and $t_j$, a simpler alternative method can be followed \cite{fang2018symmetric}. The computational details are given in \cref{app2}.
\end{remark}

\section{Numerical examples}
\label{sec5}
On the basis of aforementioned fundamentals, four examples have been presented here, some of which have been used by others to illustrate various features of the response. The example built upon the first one addressing a linear transformation followed by a nonlinear transformation, a Duffing oscillator and a nonlinear system with uncertain parameters under random excitation.

\subsection{Example \rom{1}: Linear Transformation}
\label{sec51}
Let us consider a linear transformation like:

\begin{equation}
    \label{eq: sec5_1}
\begin{aligned} x_{1} &=F_{1}(\boldsymbol{\theta})=3 \theta_{1}+5 \theta_{2} \\ x_{2} &=F_{2}(\boldsymbol{\theta})=\theta_{1}+2 \theta_{2} \end{aligned}
\end{equation}

where $\boldsymbol{\theta} \sim \mathcal{N}(\boldsymbol{\theta})$ with mean $\mu = \left[\begin{array}{l}0 \\ 0\end{array}\right]$ and $\Sigma=\left[\begin{array}{ll}1 & 0 \\ 0 & 1\end{array}\right]$. This linear transformation converts an uncorrelated Gaussian distribution into a correlated one. The inverse of \cref{eq: sec5_1} gives:

\begin{equation}
    \label{eq: sec5_2}
\begin{aligned} \theta_{1} &=2 x_1 - 5 x_2 \\ \theta_{2} &=- x_1 + 3 x_2 \end{aligned}
\end{equation}

The Jacobian of the transformation is $|J|=\left[\begin{array}{ll}2 & -5 \\ -1 & 3\end{array}\right]$ and the evolutionary PDF is:

\begin{equation}
    \label{eq: sec5_3}
    p (\boldsymbol{X}) = |J| p (\boldsymbol{\theta}) = \frac{1}{2 \pi} e^{-\frac{1}{2}(\theta_1^2+\theta_2^2)} = \frac{1}{2 \pi} e^{-\frac{1}{2}\left[\left(2 x_{1}-5 x_{2}\right)^{2}+\left(-x_{1}+3 x_{2}\right)^{2}\right]}
\end{equation}

\begin{figure}[h!]
    \centering
    \includegraphics[width=1.0\textwidth]{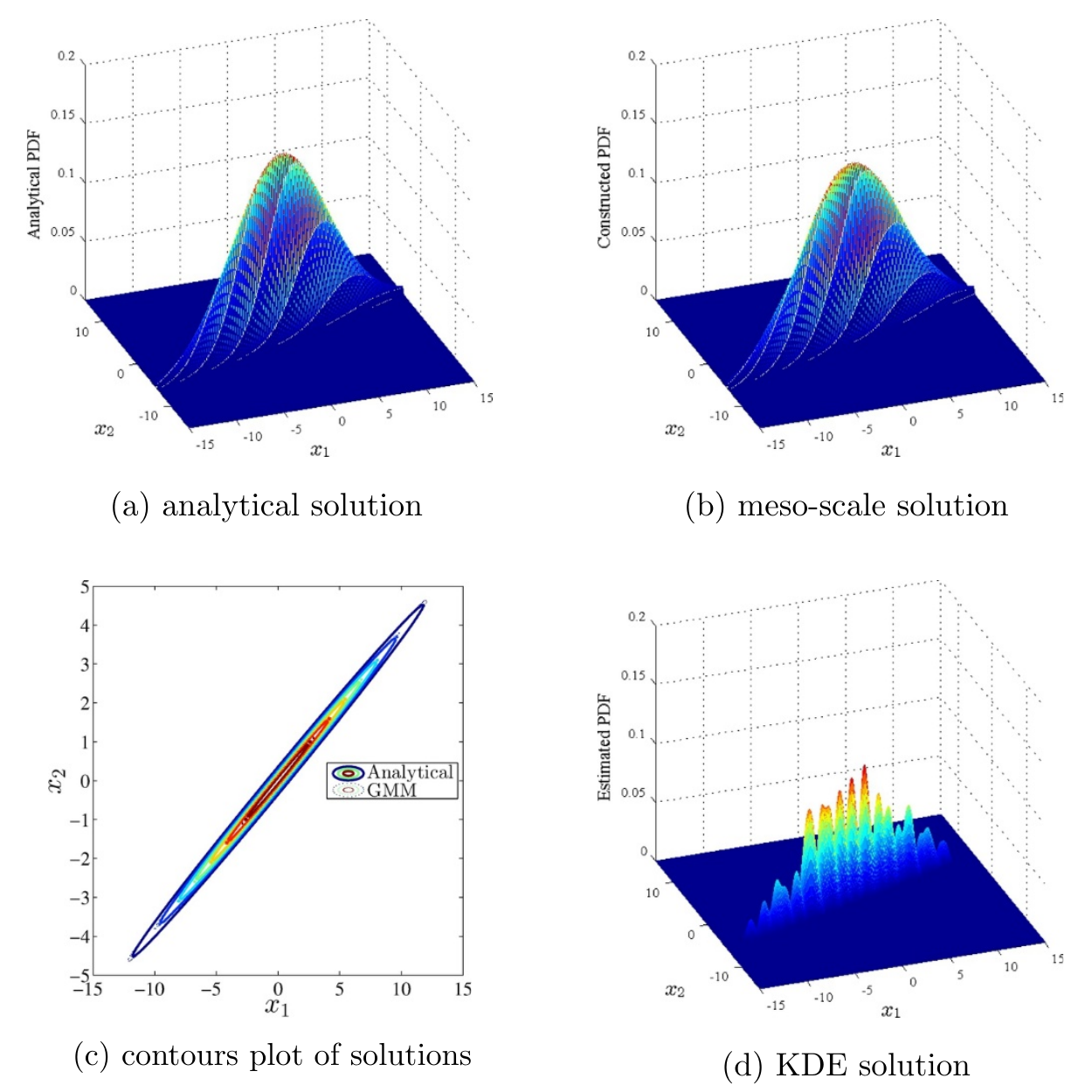}
    \caption{Example \rom{1}: summary of results.}
    \label{fig: f6}
\end{figure}

Here we use a Good lattice points (GLP) set with 89 points to determine the mean of the GMM for $p (\boldsymbol{\theta})$. Each evolutionary kernel is calculated via a third-order Gaussian cubature rule.  The analytical and meso-scale solutions of $p (\boldsymbol{X})$ are shown in \cref{fig: f6}.(a) and (b) respectively. \cref{fig: f6}.(c) plots them together to give a better comparison. In addition, the Kernel density estimation (KDE) solution for $p (\boldsymbol{X})$ based on quasi monte carlo (QMC) with the same GLP set is also calculated, as shown in \cref{fig: f6}.(d). The KDE solution is obtained by minimizing MSE with Gaussian kernel density functions. Note that the KDE solution shows a false multi-modal nature. A possible reason is that this micro-scale method loses partial information of $p (\boldsymbol{\theta})$ around each particular point of the GLP set, and fails to record its evolution, which is thus not exactly reflected in the KDE solution. By contrast, the proposed meso-scale method maintains the information on the PDF and can record the evolution precisely.

\subsection{Example \rom{2}: Nonlinear Transformation}
\label{sec52}
In this example, we consider the nonlinear transformation:

\begin{equation}
    \label{eq: sec5_4}
\begin{aligned} x_{1} &=F_{1}(\boldsymbol{\theta})=\sqrt{\theta_1^2+\theta_2^2} \\ x_{2} &=F_{2}(\boldsymbol{\theta})=\theta_1 \end{aligned}
\end{equation}

where the input uncertainties $\boldsymbol{\theta} = [\theta_1, \theta_2]$ take the same distribution stated in the previous example. By inverting \cref{eq: sec5_4} two real roots for $|x_2| < |x_1|$ can be found, as:

\begin{equation}
    \label{eq: sec5_5}
\begin{aligned} \theta_{1} &=x_1 \\ \theta_{2} & = \sqrt{x_1^2 + x_2^2} \end{aligned}
\end{equation}

The Jacobian $|J|$ in this case has the same value for both roots:

\begin{equation}
    \label{eq: sec5_6}
    |J| = \frac{x_1}{\sqrt{x_1^2 - x_2^2}}
\end{equation}

Therefore, the nonlinearly transformed probability distribution can be expressed as:

\begin{equation}
    \label{eq: sec5_7}
    p(\boldsymbol{x}) = |J| p(\boldsymbol{\theta}) = \begin{cases}
        \frac{1}{\pi} \frac{x_{1}}{\sqrt{x_{1}^{2}-x_{2}^{2}}} e^{-\frac{1}{2} x_{1}^{2}} & \text{if $x_{1}>0$ and $\left|x_{2}\right|<x_{1}$} \\
        0 & \text{else}
\end{cases}
\end{equation}

\begin{figure}[H]
    \centering
    \includegraphics[width=1.0\textwidth]{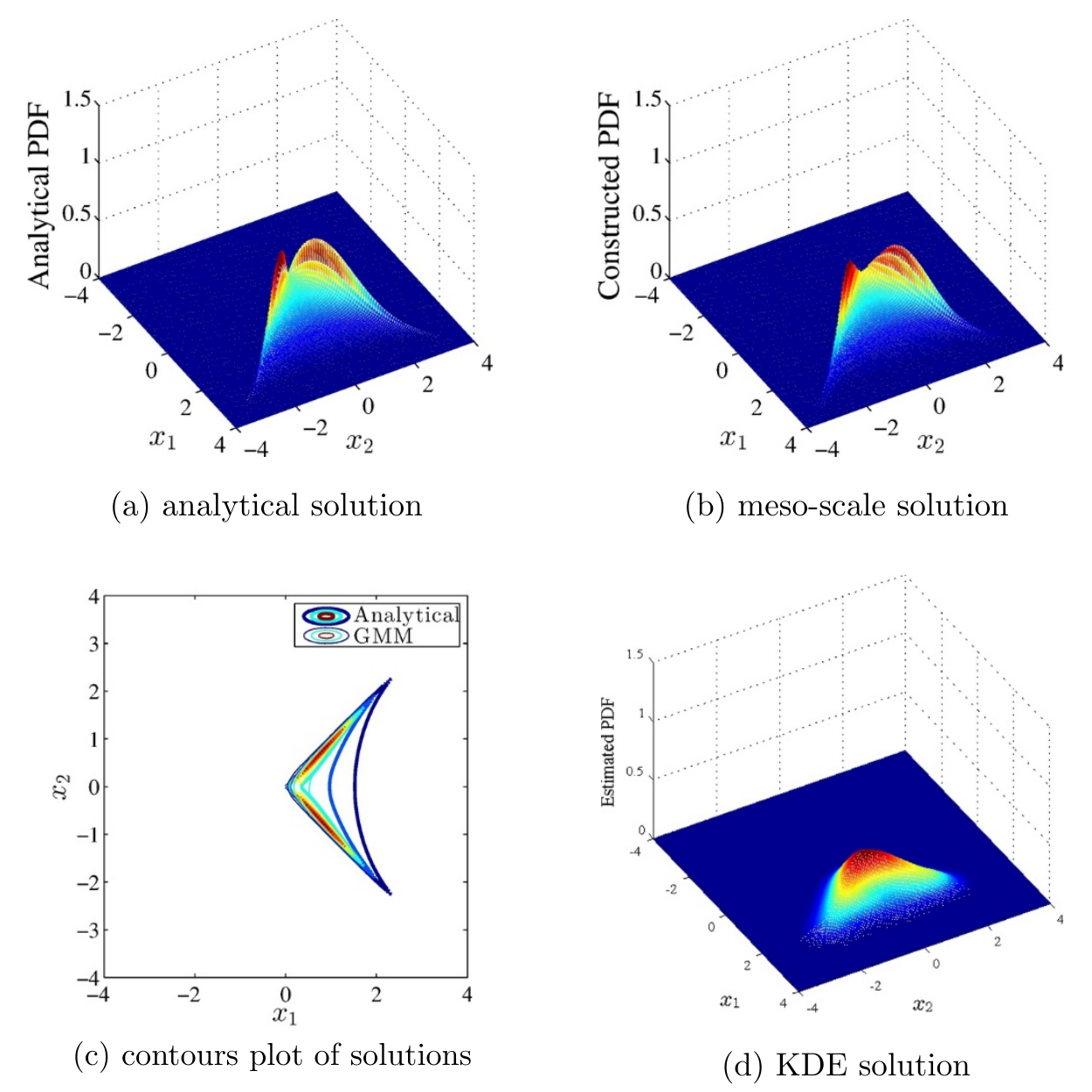}
    \caption{Example \rom{2}: summary of results.}
    \label{fig: f7}
\end{figure}

Similarly, GLP set and third-order Gaussian cubature rule are used to determine $p(\boldsymbol{x})$. The analytical and meso-scale solutions are shown in \cref{fig: f7}.(a) and (b) respectively, where the size of the mesh is $0.05 \times 0.05$. \cref{fig: f7}.(c) gives a contour plot and \cref{fig: f7}.(d) shows a KDE solution, where the bandwidth of the kernel-smoothing window is set to $0.8$. In can be observed that a false single-mode PDF rather than the exact double-mode one is computed via the normal Kernel smoothing function while meso-scale method have successfully identified the existence of two extreme regions.

\subsection{Example \rom{3}: Duffing Oscillator}
\label{sec53}
Consider a Duffing oscillator subjected to a Gaussian white noise excitation, governed by equation \cite{caughey1971nonlinear, bergman1992robust}:

\begin{equation}
    \label{eq: sec5_8}
    \begin{array}{c}\dot{X}_{1}=X_{2} \\ \dot{X}_{2}=W(t)-2 \zeta \omega_{0} X_{2}-\omega_{0}^{2} \gamma X_{1}-\omega_{0}^{2} \epsilon X_{1}^{3}\end{array}
\end{equation}

where the nominal parameters are $\zeta=0.2$, $\omega_{0}=1.0$, $\epsilon=0.10$, and $\gamma=-1.0$. The initial condition $\boldsymbol{X}_0$ is a Gaussian random vector with mean and covariance:

\begin{equation}
    \label{eq: sec5_9}
\mu\left(\boldsymbol{X}_{0}\right)=\left[\begin{array}{l}0 \\ 0\end{array}\right] ; \Sigma\left(\boldsymbol{X}_{0}\right)=\left[\begin{array}{cc}0.5 & 0 \\ 0 & 0.5\end{array}\right]
\end{equation}

The analytical solution of the stationary PDF is given by:

\begin{equation}
    \label{eq: sec5_10}
    p(x)=\frac{1}{47.9724 \sqrt{2 \pi}} \mathrm{e}^{\frac{x_{1}^{2}}{2}-\frac{x_{1}^{4}}{40}-\frac{x_{2}^{2}}{2}}
\end{equation}

\begin{figure}[h!]
    \centering
    \includegraphics[width=1.0\textwidth]{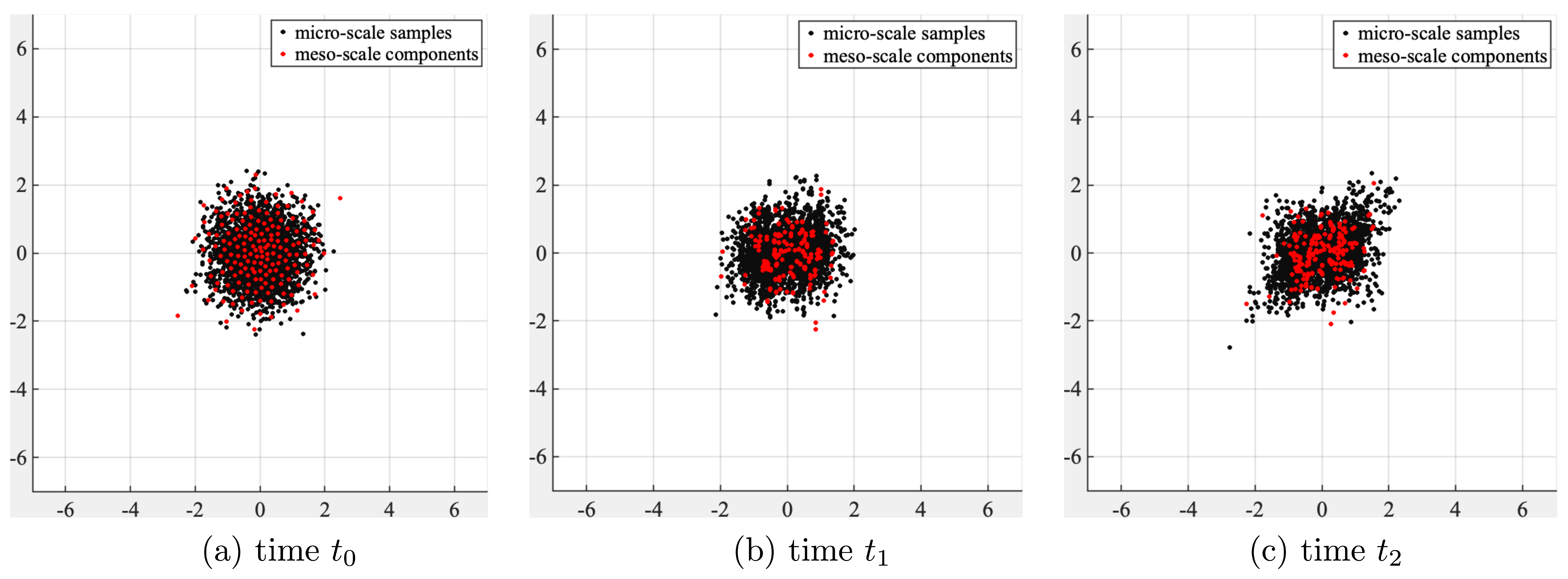}
    \caption{Probability density function evolution results.}
    \label{fig: f8}
\end{figure}

This example has been studied in the past using finite element method, which showed a distinct advantage over Monte Carlo simulation in problems of small dimension like this one. Here this problem has been revisited using the meso-scale based methodology. The initial PDF $p(\boldsymbol{X}_{t_0}) = p(\boldsymbol{X}_{0})$ is constructed by GMM with 350 component PDFs, and then updated at each time step (0.015 seconds). \cref{fig: f8} shows the sampling trajectories of the PDF from a meso-/micro-scale perspective. The computational algorithm provided in the \cref{app2} is used to calculate the influence from $W(t)$. Therefore, the total number of simulations is $4 \time 350 = 1400$, where 4 is the number of cubature points for each evolutionary kernel (See \cref{fig: f5} for details). The total number is lower than the usual number of nodes in a finite element method.

\begin{figure}[h!]
    \centering
    \includegraphics[width=1.0\textwidth]{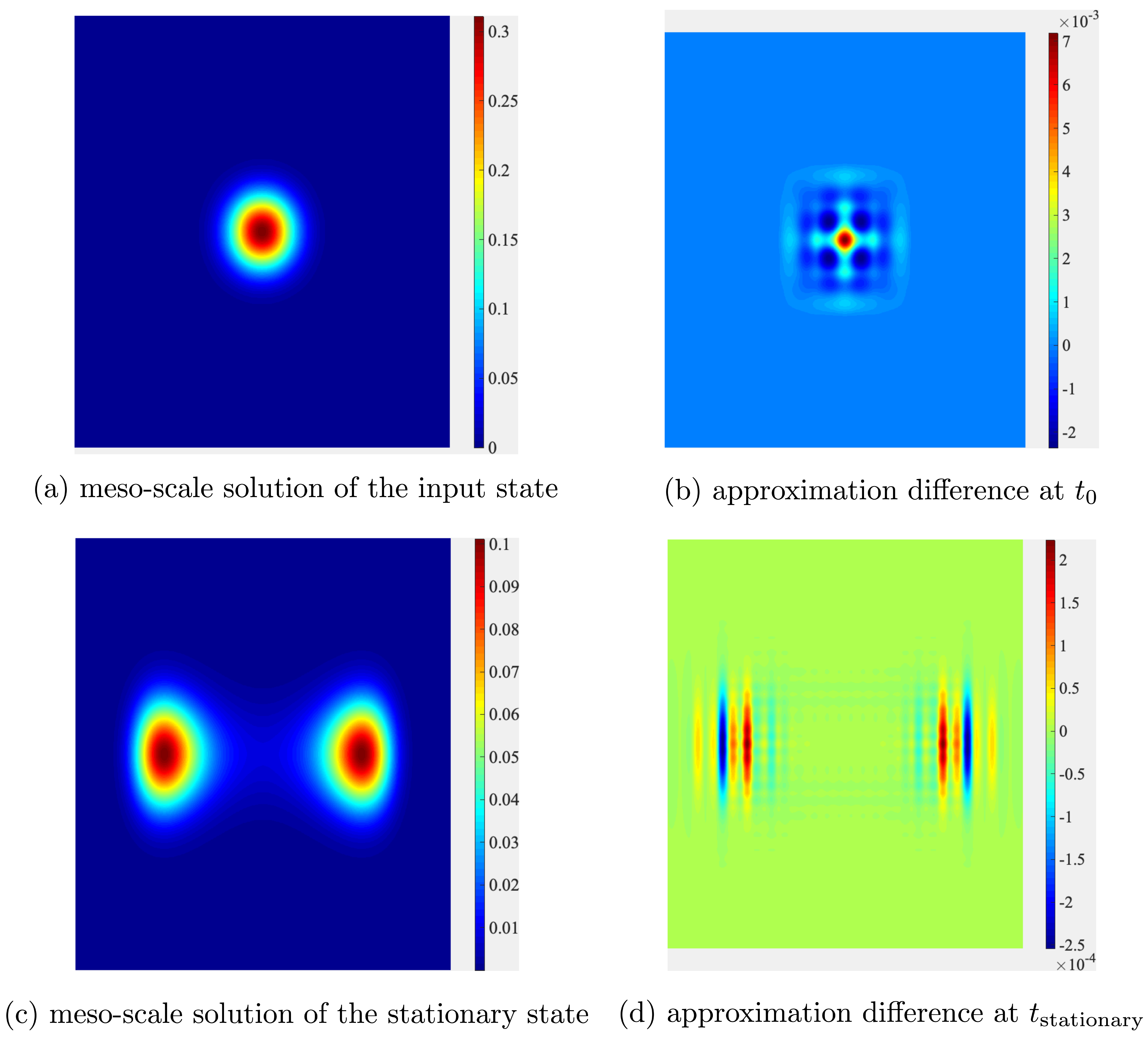}
    \caption{Meso-scale approximations of the input and stationary PDF.}
    \label{fig: f9}
\end{figure}

As time progresses, $p(\boldsymbol{X}_t$ approaches a stationary PDF, whose meso-scale solution is depicted in \cref{fig: f9}. (c). Meanwhile, the meso-scale approximation of the input uncertainties are described \cref{fig: f9}. (a). Differences between the analytical expression and the meso-scale solutions are provided for comparison. The results suggest a good match between the two solutions. Furthermore, it can be observed that the error distribution reflects the locations of the adopted meso-scale components. The maximum error is in a relatively small range, demonstrating the effectiveness of the proposed PDF evolution modeling scheme.

\subsection{Example \rom{4}: Nonlinear Structure with Uncertain Parameters Subjected to Random Excitation}
\label{sec54}
In the last example, let us consider a 10-story 2-bay uncertain shear frame subjected to random ground motion \cite{li2009stochastic}. The lumped masses, $m_1, m_2, \dots, m_{10}$, are listed in \cref{table: t1}:

\begin{table}[H]
    \centering
    \begin{tabular}{l l l l l l l l l l}
    \hline
    $\boldsymbol{m}_1$ & $\boldsymbol{m}_2$ & $\boldsymbol{m}_3$ & $\boldsymbol{m}_4$ & $\boldsymbol{m}_5$ & $\boldsymbol{m}_6$ & $\boldsymbol{m}_7$ & $\boldsymbol{m}_8$ & $\boldsymbol{m}_9$ & $\boldsymbol{m}_{10}$ \\
    \hline
    0.5 & 1.1 & 1.1 & 1 & 1 & 1.1 & 1.3 & 1.2 & 1.2 & 1.2 \\
    \hline
    \end{tabular}
    \caption{The lumped masses of the structure ($\times 10^5 kg$).}
     \label{table: t1}
\end{table}

Meanwhile, other structural characteristics are given as: $h = 4 m, h^{'} = 3 m$; column square cross-section with dimension $S = 500 mm \times 500 mm$; beams with infinite stiffness. The damping matrix is $C = a M + b K$, where $M$ and $K$ are respectively the mass matrix and the initial stiffness matrix. Assume $a = 0.01$ and $b = 0.005$. The Bouc-Wen model is used to model the restoring force as:

\begin{equation}
    \label{eq: sec5_11}
    R_{T}(x, z)=a k x+(1-\alpha) k z
\end{equation}

where $k$ is the initial stiffness; $x$ is the inter-story drift and $z$ is the hysteretic component satisfying:

\begin{equation}
    \label{eq: sec5_12}
    \dot{z}=A \dot{x}-\beta|\dot{x}||z|^{n-1} z-\gamma \dot{x}|z|^{n}
\end{equation}

in which the parameters take the value $\alpha = 0.01$, $A = 1.2$, $\beta = 1.4$, $\gamma = 0.2$, and $n = 1$. The initial Young’s modulus $E$ is an uncertain parameter. The random ground motion is represented by a randomly scaled El Centro record with peak ground acceleration (PGA) as the random variable. The total probabilistic information is listed in \cref{table: t2}.

\begin{table}[H]
    \centering
    \begin{tabular}{l l l l}
    \hline
    \textbf{Parameter} & \textbf{Distribution} & \textbf{Mean} & \textbf{C.O.V.} \\
    \hline
    E & Normal & $3.0 \times 10^{10}$ Pa  & $0.1$ \\
    PGA & Normal & $2.0 m/s^2$ & $0.1$ \\
    \hline
    \end{tabular}
    \caption{The probabilistic information of the random parameters.}
     \label{table: t2}
\end{table}

\begin{figure}[H]
    \centering
    \includegraphics[width=1.0\textwidth]{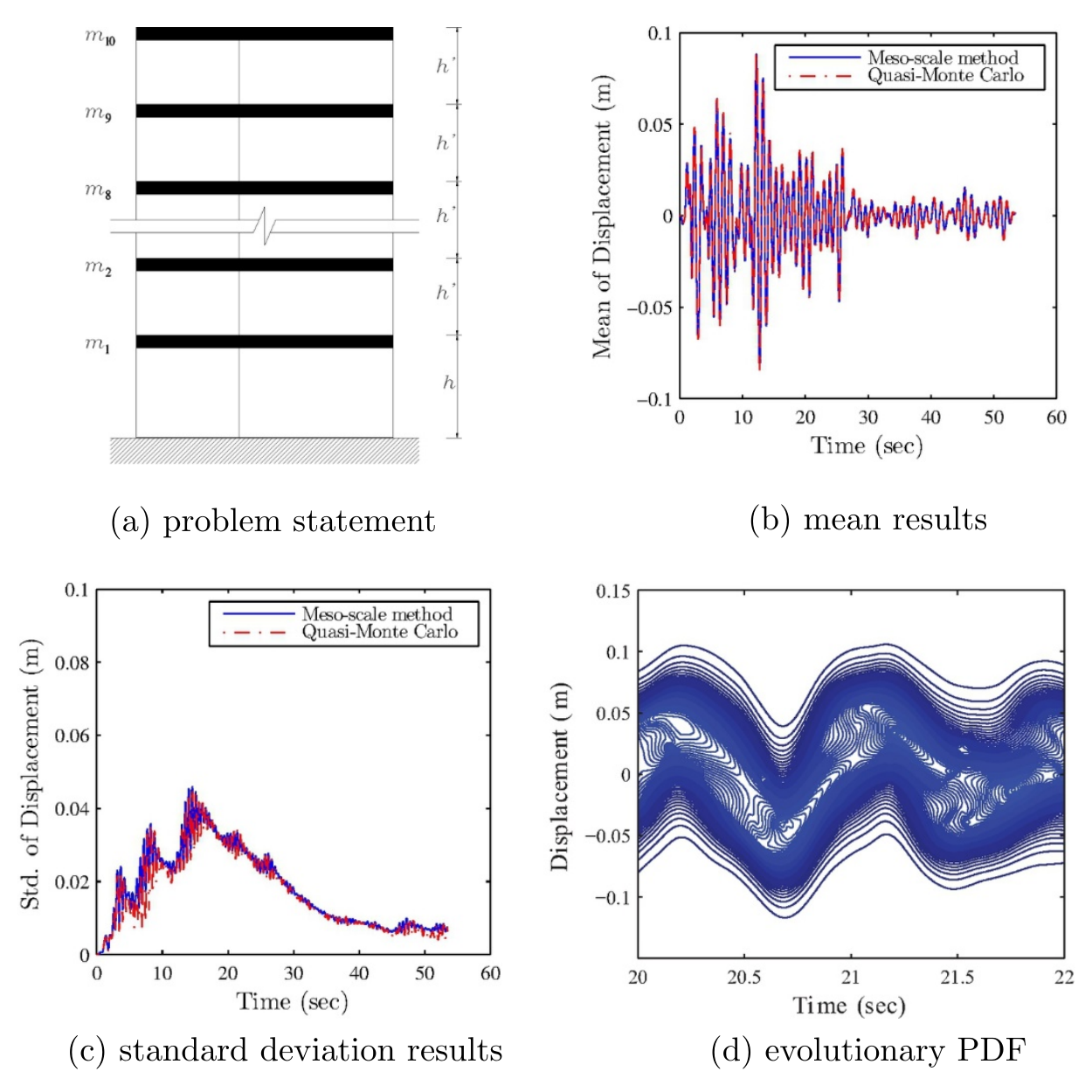}
    \caption{Example \rom{4}: summary of results.}
    \label{fig: f10}
\end{figure}

As previously discussed in \cref{sec41}, this system can be recast into a conservative model by placing the total random variables in an augmented random initial condition. For comparison, the meso-scale method and QMC are used to compute the probability evolution. The second-order statistics of the top floor displacement calculated by both methods are shown in \cref{fig: f10}.(b) and (c). Note that both methods give similar second-order statistics. In addition, the meso-scale method can also provide the evolution of probability as shown in \cref{fig: f10}.(d), which is not given by QMC.

\section{Concluding remarks}
\label{sec6}
In this paper, the probability evolution has been investigated from a meso-scale perspective. The proposed scheme is able not only to maintain the information concerning the PDF with an equivalent statistical structure but also to track the evolution of this statistical structure. This is accomplished by utilizing a Gaussian mixture model (GMM) for the representation of the input PDF, and solving a series of mixtures of Gaussian integrals for the probability evolution. We have demonstrated that such meso-scale computation frees the macro-scale methods from the limitations of the second-order statistics and enhancing the expressibility of micro-scale methods by maintaining the PDFs around samples. Furthermore, the connection between the proposed meso-scale scheme to the standard conservative and Markov models has been established in the context of stochastic modeling. The third-degree spherical-radial cubature rule is introduced to further reduce the number of parameters that are involved in the optimization of GMM. The efficacy of the meso-scale method is verified by several examples. In summary, the provided results have demonstrated the merit of the proposed method and a new meso-scale perspective regarding examining the probability evolution it offers.

For the further extensions to this work, future studies may address issues as: (1) generalizing the method to the probabilistic regime, i.e. incorporating the Bayesian framework into the current scheme; (2) scaling the uncertainty quantification process to an ultra high-dimensional stochastic process via the adoption of deep latent space model, variational autoencoder (VAE), generative adversarial networks (GANs), etc; and (3) applying the method to a broader range of different types of engineering problems.

\section*{Acknowledgement}
This research has been supported in part by the National Science Foundation under Grant Agreement No. 1520817 and No. 1612843. A. Kareem gratefully acknowledges the financial support of Robert M Moran professorship.

\bibliographystyle{elsarticle-num}
\bibliography{sample}

\appendix
\section{Expectation-maximization algorithm for covariances computation}
\label{app1}
Expectation-maximization (EM) algorithm is closely related to the maximum likelihood estimation (MLE) that is an effective approach that underlies many machine learning algorithms. The EM algorithm performs MLE for learning parameters in probabilistic models with latent variables. With determined $\{\omega_k \}_{i=k}^{k}$ and $\{ \boldsymbol{\mu}_k \}_{i=k}^{k}$ (See \cref{sec32}), using EM algorithm to optimize $\{ \boldsymbol{\Sigma}_k \}_{i=k}^{k}$ can be summarized to six steps:

\begin{itemize}
    \item[1] Initialize the set $\{ \hat{\boldsymbol{\mu}_{k}} \}_{k=1}^{K}$ as $\mathcal{P}$. 
    \begin{itemize}
        \item[1.1] If $\mathcal{P}$ is an LDS, then randomly choose an initial set $\{ \boldsymbol{\Sigma}_{k} \}_{k=1}^{K}$ and an auxiliary point set $\{ \boldsymbol{\theta}_{j}^{'} \}_{j=1}^{M}$, where $M \gg K$.
        \item[1.2] $\mathcal{P}$ is a cluster set, then set the initial value of $\boldsymbol{\Sigma}_{k}$ as:
        \begin{equation}
            \label{eq: app1_1}
            \boldsymbol{\Sigma}_{k} = \frac{1}{N_{i}} \sum_{j=1}^{N_i} (\boldsymbol{\theta}_{(k,j)} - \boldsymbol{\mu}_{k}) (\boldsymbol{\theta}_{(k,j)} - \boldsymbol{\mu}_{k})^T
        \end{equation}
        where $\{ \boldsymbol{\theta}_{(k,j)} \}_{j=1}^{N_i}$ is the auxiliary point set assigned to the $k^{th}$ cluster, with $\sum_{k=1}^{N} N_{k}=M$.
    \end{itemize}
    \item[2] Estimate the log likelihood:
    \begin{equation}
        \label{eq: app1_2}
        \ln p \left( \boldsymbol{\theta}_{k} | \boldsymbol{\Sigma}_{\theta_k} \right) = \sum_{k=1}^{K} \ln \left[ \sum_{j=1}^{M} \frac{1}{K} \mathcal{N}_{k} \left( \theta_j^{\prime} | \Sigma_{\theta_{k}} \right) \right]
    \end{equation} 
    \item[3] Evaluate the responsibilities (expectation step):
    \begin{equation}
        \label{eq: app1_3}
        \lambda_{(k,j)} = \frac{\mathcal{N}_{k} \left( \theta_j^{\prime} | \Sigma_{\theta_{k}} \right)}{\sum_{k=1}^{K} \mathcal{N}_{k} \left( \theta_j^{\prime} | \Sigma_{\theta_{k}} \right)}
    \end{equation} 
    \item[4] Re-estimate the following parameters (maximization step):
    \begin{equation}
        \label{eq: app1_4}
        \Sigma_{\theta}^{(k)} = \frac{1}{\beta^{k}} \sum_{j=1}^{M} \lambda_{(k,j)} (\boldsymbol{\theta}_{(k,j)}^{'} - \boldsymbol{\mu}_{k}) (\boldsymbol{\theta}_{(k,j)}^{'} - \boldsymbol{\mu}_{k})^T
    \end{equation} 
    where $\beta^{k} = \sum_{j=1}^{M} \lambda_{(k,j)}$. Then re-estimate the log likelihood with the $\Sigma_{\theta}^{(k)}$ given by \cref{eq: app1_4}.
    \item[5] Check for convergence for either the parameters or the log likelihood.
    \item[6] If not converged, repeat steps $2 \sim 5$; if converged, set $\left\{\hat{\Sigma}_{\theta}^{(k)}\right\}_{k=1}^{K}=\left\{\Sigma_{\theta}^{(k)}\right\}_{k=1}^{K}$.
\end{itemize}

\section{Computational procedures for Markov models}
\label{app2}
The overall updating scheme for computing the first two statistical moments can be summarized to two steps.

\begin{itemize}
    \item[1] Rewrite the mean of the $i^{th}$ evolutionary kernel (\cref{eq: sec4_15}) as:
    \begin{equation}
        \label{eq: app2_1}
        \begin{aligned}
        \mu^{(i)}\left(X_{t_{j}}\right) & \approx \frac{1}{N^(i)} \sum_{k=1}^{N^(i)} \left[x_{t_{j-1}}^{(i, k)}+\int_{t_{j-1}}^{t_{j}} G\left(x_{s}, s\right) \mathrm{d} s+\int_{t_{j-1}}^{t_{j}} A\left(x_{s}, s\right) \mathrm{d} B_{s}\right] \\
        & = \frac{1}{N^(i)} \sum_{k=1}^{N^(i)} \left[x_{t_{j-1}}^{(i, k)}+\int_{t_{j-1}}^{t_{j}} G\left(x_{s}, s\right) \mathrm{d} s\right]
        \end{aligned}
    \end{equation}
    and $L_{t_{j}}^{(i, k)}$ (\cref{eq: sec4_17}) as:
    \begin{equation}
        \label{eq: app2_2}
        L_{t_{j}}^{(i, k)}=\left[x_{t_{j-1}}^{(i, k)}+\int_{t_{j-1}}^{t_{j}} G\left(x_{s}, s\right) \mathrm{d} s-\mu^{(i)}\left(X_{t_{j}}\right)\right]+\int_{t_{j-1}}^{t_{j}} A\left(x_{s}, s\right) \mathrm{d} B_{s}^{(i, k)}
    \end{equation}
    Then, we have:
    \begin{equation}
        \label{eq: app2_3}
        \sum^{(i)}\left(X_{t_{j}}\right) \approx \frac{1}{N^{(i)}} \sum_{k=1}^{N^{(i)}}\left[L_{x, t_{j}}^{(i, k)} L_{x, t_{j}}^{(i, k)^{T}}+L_{B, t_{j}}^{(i, k)} L_{B, t_{j}}^{(i, k)^{T}}\right]
    \end{equation}
    where
    \begin{equation}
        \label{eq: app2_4}
        L_{x, t_{j}}^{(i, k)}=x_{t_{j-1}}^{(i, k)}+\int_{t_{j-1}}^{t_{j}} G\left(x_{s}, s\right) \mathrm{d} s-\mu^{(i)}\left(X_{t_{j}}\right)
    \end{equation}
    and
    \begin{equation}
        \label{eq: app2_5}
        L_{B, t_{j}}^{(i, k)}=\int_{t_{j-1}}^{t_{j}} A\left(x_{s}, s\right) \mathrm{d} B_{s}^{(i, k)}
    \end{equation} 
    The above equations indicate that $W_{t_{j-1}:t_j}$ affects the covariance of each evolutionary kernel without causing any influence to its mean.   
    \item[2] Utilize the samples of the additive noise to estimate the updated values of the mean and covariance of each evolutionary kernel:
    \begin{equation}
        \label{eq: app2_6}
        \mu^{(i)}\left(\boldsymbol{X}_{t_{j}}\right) \approx \frac{1}{N^{(i)}} \sum_{k=1}^{N^{(i)}}\left\{\frac{1}{N_{B}} \sum_{l=1}^{N_{B}}\left[x_{t_{j-1}}^{(i, k)}+\int_{t_{j-1}}^{t_{j}} G\left(\boldsymbol{x}_{s}, s\right) \mathrm{d} s+\int_{t_{j-1}}^{t_{j}} A\left(\boldsymbol{x}_{s}, s\right) \mathrm{d} B_{s}^{(l)}\right]\right\}
    \end{equation}
    and
    \begin{equation}
        \label{eq: app2_7}
        L_{t_{j}}^{(i, k)}=\frac{1}{N_{B}} \sum_{l=1}^{N_{B}}\left[x_{t_{j-1}}^{(i, k)}+\int_{t_{j-1}}^{t_{j}} G\left(x_{s}, s\right) \mathrm{d} s+\int_{t_{j-1}}^{t_{j}} A\left(x_{s}, s\right) \mathrm{d} B_{s}^{(l)}-\mu^{(i)}\left(\boldsymbol{X}_{t_{j}}\right)\right]
    \end{equation}
    where $N_B$ represents the number of samples of the additive noise. This estimation can be made through the Monte Carlo simulation. Since only $\int_{t_{j-1}}^{t_{j}} A\left(x_{s}, s\right) \mathrm{d} B_{s}^{(l)}$ needs to be calculated, this would not require a significant computational effort.
\end{itemize}

\end{document}